# Climate-dependent enhancement of radiative cooling with mirror structures

**Jaesuk Hwang**[a,*]

[a] Center for Quantum Technologies, National University of Singapore, 3 Science Drive 2, 117543, Singapore

**Abstract**. Radiative cooling exploits the imbalance between the thermal emission from the radiative cooling surface and the downward atmospheric emission. Since the atmospheric emission power is polar angle-dependent, a mirror structure can be used to increase this imbalance and to amplify the net cooling power. The degree of amplification is determined by various parameters such as the sky emissivity, the geometry of the mirror structure and the degree of thermal insulation. A parametric study of the aperture mirror-enhanced radiative cooling is presented using a model atmosphere, characterized by an average sky window emissivity and the ambient temperature. A counterintuitive finding is obtained, namely that the aperture mirror structure is more effective in the tropics than in the desert, both in terms of the cooling power and the temperature reduction. The power enhancement obtainable from a relatively simple mirror structure can be significant. For example, in the tropics, the cooling power can be enhanced by more than 40%. The aperture mirror structure holds potential to be a practical augmentation to improve the stagnant temperature and the response time of radiative cooling devices.

*E-mail: jhwang@nus.edu.sg

## 1 Introduction

The energy demand for cooling increases as the temperature on the earth constantly rises due to the greenhouse effect[1-3]. The use of compression-based systems, such as air conditioners, accounts for an estimated 20% of electricity consumption in buildings globally[4]. However, the compression-based systems merely move heat from one place to another, causing environmental problems such as the urban heat island (UHI)[5] and more greenhouse gas emissions, only to accelerate the global warming. As a zero-energy, passive cooling solution which does not release heat to the environment, radiative cooling[6-20] is considered promising. When objects on the earth, around 300K temperature, emit towards the extreme coldness of the outer space, at 3 K temperature, the object can cool below the ambient temperature passively without any energy input[21, 22]. Radiative cooling devices offer a distinct advantage of dissipating heat to the sky, not to the environment, and the capability to provide a cooling power at a sub-ambient temperature, albeit a passive cooling mechanism.

The outer space temperature at 3 K temperature is effectively screened from the terrestrial level by the atmosphere acting as a greenhouse, therefore not directly accessible as a heat sink for radiative cooling. The thermal radiation from the earth can escape the atmosphere because within the wavelength band between 8 to 14 microns wavelength, the atmosphere is less emissive and absorptive than the blackbody at ambient temperature. This wavelength band is called the sky window (often a smaller wavelength band is used, such as 7.9 to 13 microns)[10, 12]. Due to the sky window, the downward thermal radiation, so-called downwelling, is out of thermal equilibrium and does not follow the Planck spectrum. As a result, the sky usually appears colder than the atmosphere, although at a temperature higher than 3K. Radiative cooling exploits the imbalance in the thermal radiation in the sky window. Using a substrate absorbing and emitting efficiently in the sky window, heat can be pumped radiatively towards the outer space through the atmosphere. The downwelling within the sky window wavelength range is polar angle-dependent, weakest in the zenith direction and strongest in the direction close to the horizontal[23]. Based on this property of the atmosphere, the imbalance in the thermal radiation can be increased using a heat mirror structure and the radiative cooling power can be amplified[6]. The heat mirror structure can direct the emission from the radiative cooling surface closer to the zenith direction and block the downwelling at an oblique angle from being incident on the radiative cooling surface. For example, planar mirrors were disposed near an upright radiative cooling surface to guide the thermal radiation upwards and to exploit radiative cooling of both sides of the radiative cooling surface[24]. A reflecting parabolic trough structure was placed around a pipe with a radiative cooling coating to reject the thermal radiation from the oblique angle and the ground[25, 26]. For a planar radiative cooling surface facing the zenith direction, an aperture mirror structure was used in the shape of a tapered waveguide[24], a truncated cone[16, 27-30] and a parabola[31], reminiscent of the concentrating mirror structures from the solar thermal harvesting[32].

This report explores the aperture mirror structure disposed around a radiative cooling surface. Since the enhancement originates solely from the optical arrangement around existing radiative cooling surface, the use of a heat mirror structure has a significant engineering implication and offers practical advantages in thermal management exploiting radiative cooling. The cooling power from the radiative cooling surface scales with the surface area directly facing the sky and the aperture mirror structure can provide a higher cooling power from the fixed surface area. A higher cooling power would lower the reachable steady-state stagnant temperature and decrease the time to cool down a given thermal mass, thereby offering a broader range of option for the design of a passive cooling system in terms of the target temperature and the time constant. Although the mirror structure increases the volume occupied by the radiative cooling device for a given area, the added volume is shielded from the movement of air and therefore works to mitigate the convective heat gain and can serve as a structure to support additional convective cover[33]. In case the mirror structures are deployed on the roof[34], the restriction in space in the vertical direction is of relatively less concern. Also, in an urban area with densely packed building structures with different heights, the top surface of each building may not have a full hemispherical access of the sky. This aperture mirror structure addresses this issue by reimaging the view from the radiative cooling surface on the roof to a small solid angle near the zenith direction. Therefore, the enhancement of radiative cooling power using heat mirror structures hold promise in the passive thermal management system aided by radiative cooling.

The enhancement from using the aperture mirror structure results from the interplay of the geometry of the aperture mirror structure, the emissivity of the radiative cooling surface, $e_s(\eta, \lambda)$ and the emissivity of the atmosphere $e_a(\eta,\lambda)$, where $\eta$ is the polar angle from the zenith direction and $\lambda$ is the wavelength. The emissivity spectrum of the radiative cooling surface $e_s(\eta,\lambda)$ can be measured and the aperture mirror can be analysed using ray tracing[28, 30]. Obtaining the atmospheric emissivity spectrum $e_a(\eta,\lambda)$ calls for a practical strategy, because it depends on the local climate, the exact time of the day and the weather condition[16, 35-39]. Although the full spectrum of the atmospheric emissivity can be measured, a constant monitoring is often not viable. Since the relationship between the hemispherical average value of the sky emissivity and the dew point temperature is empirically known and an interpolated curve is available[40], the average atmospheric emissivity can be deduced from the dew point temperature measured at each time point. The full atmospheric emissivity spectrum can be simulated[16, 35-38] using the MODTRAN programs[41, 42] but the simulation results may only provide a reasonable estimate for a stationary and uniform sky. The use of a model atmosphere, characterised by the average sky window emissivity and the ambient temperature, can be a practical alternative. The use of the model atmosphere can provide immediate insights on the dependence on the sky conditions without the need for an extensive simulation[43].

In this report, a detailed parametric study of aperture mirror-enhanced radiative cooling system is presented. The analysis of Smith[43] is revisited and augmented by including the 'reciprocal ray' contribution[30], which can be significant especially when the radiative cooling surface is a blackbody emitter. The resultant formulation will be applied to the two extreme cases, a blackbody emitter and a selective emitter, and one realistic emitter example, the 3M specular reflector. The dependence of the aperture mirror enhancement on the regionality, the mirror geometry, the degree of thermal insulation will be investigated in detail.

## 2 Modelling atmospheric emissivity

A simple model can be used to characterise the atmospheric spectrum dividing the wavelength range into two sections: outside the sky window, the atmosphere is a blackbody at the ambient temperature and throughout the sky window, the atmosphere is taken to be a gray body with a single average value of emissivity[11, 43, 44]. Within the sky window, a net outgoing thermal radiation results, whereas outside the sky window, the outgoing thermal radiation and the downwelling mostly balance out. According to this model, the emissivity of the atmosphere $e_a(\eta, \lambda)$ is written as follows:

$$e_{a,nsw}(\eta, \lambda) = 1 \text{ outside the sky window,}$$

$$e_{a,sw}(\eta) \approx 1 - (1 - e_{avg,sw})^{1/\cos\eta} \text{ within the sky window,} \quad (1)$$

where $e_{avg,sw}$ is the average sky window emissivity in the zenith direction, $\eta = 0$. The resulting atmospheric spectrum, both absorption and downward emission, is given by as $e_a(\eta, \lambda) P(\lambda, T_a)$, where $P(\lambda, T_a)$ is the Planck spectrum at ambient temperature $T_a$. The atmospheric emission spectrum outside the sky window is approximated as the Planck radiation spectrum at ambient temperature $T_a$, $P(\lambda, T_a)$, regardless of the angle $\eta$. The advantage of this simple model is that based only on two parameters, the ambient temperature $T_a$ and the average sky window emissivity $e_{avg,sw}$, the atmospheric spectrum can be characterised in enough detail. For evaluating the net radiative cooling power, the detailed spectral features, such as the known ozone peak at 9.6 μm, are not considered, if the radiative cooling surface is emissive throughout the sky window. In relation to the polar angle dependence of the emissivity in the sky window, the empirical "cosine approximation", the angle dependence of $1/\cos\eta$ power to the transmissivity, $1 - e_{avg,sw}$, is used[23, 45]. One of the most determining factors for the sky window transparency is known to be the water vapour content in the air of the region, correlated with the local humidity[39, 43, 46]. Therefore, in this atmospheric model, the regionality and the changing weather condition are represented in the average sky window emissivity $e_{avg,sw}$. Since the average sky window emissivity $e_{avg,sw}$ can be measured directly, as will be discussed in more detail in section 8, the atmospheric model also lends itself to an alternative way of measuring sky emissivity to that using a Pyrgeometer[29, 47].

## 3    Aperture mirror structure

Figure 1 shows the basic geometry of the configuration under discussion. A planar radiative cooling surface, at temperature $T_s$, is disposed normal to the zenith direction, under ambient condition at temperature $T_a$. The aperture mirror structure is disposed around the radiative cooling surface to expose the radiative cooling surface into the zenith direction, $\eta = 0$, but to limit the solid angle of the view around the zenith direction to the maximum angle $\eta_{max}$. The aperture mirror structure redirects the thermal emission from the radiative cooling surface towards the zenith direction and blocks the oblique downwelling incident outside the angle $\eta_{max}$. The internal surface of the aperture mirror structure must be a good heat mirror, such as polished aluminium surface, to maximise the specular reflection and to minimise the diffuse scattering, which will be incident on the radiative cooling surface. It is relatively straightforward to produce by hand-polishing an aluminium surface with 95 % total reflectivity and 90 % specular reflectivity over the thermal wavelength band[29]. In this report, an ideal reflector will be assumed for simplicity, which provides an upper bound of the enhancement effect. The shape of the aperture mirror structure will not be fixed to a specific geometry but two conditions are required, namely that the downwelling incident outside the maximum angle $\eta_{max}$ is blocked from being incident on the radiative cooling surface and that the thermal emission from the radiative cooling surface is redirected the zenith direction with atmospheric emissivity $e_a(0, \lambda)$. The latter condition is strictly met when a parabolic mirror is used and the radiative cooling surface is positioned at the focus of the parabolic mirror. If the lateral extent of the radiative cooling surface exceeds the area covered by the focal volume or if other shapes of the aperture mirror structure is used, for example, a truncated cone, the rays from the radiative cooling substrate can be distributed within a finite solid angle about the zenith direction. Even in this case, it is a good approximation that the thermal emissions are redirected in the zenith direction because the emissivity changes slowly with angle $\eta$, in particular, quadratically because $e_{a,sw}(\eta) \approx e_{avg,sw} - \frac{1}{2}(1 - e_{avg,sw})\ln(1 - e_{avg,sw})\eta^2$. For example, for an opaque sky window, where the average sky window emissivity $e_{avg,sw} = 0.6$, even with a 36° spread about the zenith, the emissivity $e_a\left(\frac{\pi}{10}, \lambda\right)$ is only 5 % smaller than that in the zenith direction $e_a(0, \lambda)$.

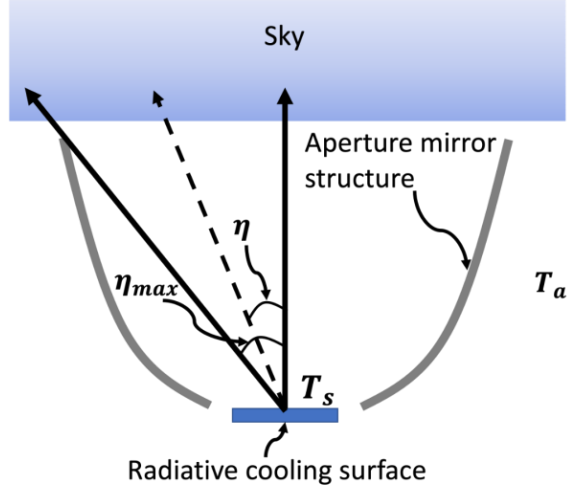

**Fig. 1** The geometry of aperture mirror structure.

## 4  Net radiative cooling power

The net radiative cooling power $P_{rad}(T_a, T_s)$ as a function of the ambient temperature $T_a$ and the radiative cooling substrate temperature $T_s$ is given by:

$$P_{rad}(T_a, T_s) = P_1(T_s) - P_2(T_a) - P_3(T_a) - P_{parasitic}(T_a, T_s). \qquad (2)$$

$P_1$ is the power radiated from the radiative cooling substrate at temperature $T_s$:

$$P_1(T_s) = \int_0^{\pi/2} d(\sin^2\eta) \int_0^\infty d\lambda P(\lambda, T_s) e_s(\eta, \lambda), \qquad (3)$$

where $e_s(\eta, \lambda)$ is the angle-dependent emissivity spectrum of the radiative cooling surface and $P(\lambda, T_s)$ is the Planck spectrum. $P_2$ is the downwelling absorbed by the radiative cooling substrate within the acceptance angle $\eta_{max}$ of the aperture mirror structure:

$$P_2(T_a) = \int_0^{\eta_{max}} d(\sin^2\eta) \int_0^\infty d\lambda P(\lambda, T_a) e_a(\eta, \lambda) e_s(\eta, \lambda)$$
$$= \int_0^{\pi/2} d(\sin^2\eta) \left( \int_{SW} d\lambda P(\lambda, T_a)(1 - (1 - e_{avg,sw})^{1/\cos\eta}) e_s(\eta, \lambda) + \int_{NSW} d\lambda P(\lambda, T_a) e_s(\eta, \lambda) \right), \quad (4)$$

where SW represents the wavelength range of the sky window and NSW represents the wavelength range outside the sky window. $P_3$ is the "reciprocal rays" term[30], which accounts for the downwelling in the zenith direction incident on the radiative cooling surface by being reflected by the aperture mirror structure and absorbed by the radiative cooling surface:

$$P_3(T_a) = \int_{\eta_{max}}^{\pi/2} d(\sin^2\eta) \int_0^\infty d\lambda P(\lambda, T_a) e_a(0, \lambda) e_s(\eta, \lambda)$$
$$= \int_{\eta_{max}}^{\pi/2} d(\sin^2\eta) \left( \int_{SW} d\lambda P(\lambda, T_a) e_{avg,sw} e_s(\eta, \lambda) + \int_{NSW} d\lambda P(\lambda, T_a) e_s(\eta, \lambda) \right). \quad (5)$$

Note that $P_3$ accounts for the downwelling which would not have been incident on the radiative cooling surface if the aperture mirror structure was absent[30]. $P_{parasitic}$ is the parasitic heat gain into the radiative cooling surface via conduction and air convection and given by $P_{parasitic}(T_a, T_s) = h_{eff}(T_a - T_s)$, where $h_{eff}$ is the heat transfer coefficient, which characterises the degree of thermal insulation.

## 5 Two extreme cases: blackbody emitter and selective emitter

As illustrative examples, two extreme cases for the radiative cooling surfaces will be considered: a blackbody emitter and a selective emitter. Both are assumed to be perfectly absorptive within the sky window, $e_s(\eta, \lambda) = 1$ as is required for strong radiative cooling. For the blackbody emitter, it is assumed that $e_s(\eta, \lambda) = 1$ independent of the angle $\eta$ and wavelength $\lambda$. For the selective emitter, it is assumed that $e_s(\eta, \lambda) = 0$ outside the sky window, independent of the angle $\eta$. When the parasitic heat gain $P_{parasitic}$ is disregarded for simplicity, equation (2) simplifies to equations (6) and (7) for the blackbody and the selective emitter cases, respectively.

$$P_{rad}^{BB}(T_a, T_s) = \sigma T_s^4 - \sigma T_a^4\{sin^2\eta_{max}(1 - \rho_{sw}(T_a)) + I_{sw}(\eta_{max})\rho_{sw}(T_a)\} - \sigma T_a^4(1 - sin^2\eta_{max})\{\rho_{sw}(T_a)e_{a,sw}(0) + (1 - \rho_{sw}(T_a))\}. \quad (6)$$

$$P_{rad}^{SEL}(T_a, T_s) = \sigma T_s^4 \rho_{sw}(T_s) - \sigma T_a^4 I_{sw}(\eta_{max})\rho_{sw}(T_a) - \sigma T_a^4(1 - sin^2\eta_{max})\rho_{sw}(T_a)e_{a,sw}(0). \quad (7)$$

$\rho_{sw}(T)$ and $I_{sw}(\eta_{max})$ are as defined in ref[43]: $\rho_{sw}(T)$ is the fraction of the total energy under the Planck spectrum at temperature $T$ within the sky window and ranges from 0.3 to 0.35 around ambient temperature. $I_{sw}(\eta_{max})$, the aperture factor, is given by $I_{sw}(\eta_{max}) = \int_0^{\eta_{max}} d(sin^2\eta)e_{a,sw}(\eta)$, which accounts for the energy of the downwelling within the sky window wavelength range received by the radiative cooling surface, without being reflected by the concentrator.

## 6 Regional dependence of enhancement of blackbody emitter and selective emitter

The average sky window emissivity $e_{avg,sw}$ depends strongly on the meteorological conditions of each geographical region, especially local humidity[43, 46]. The regional dependence of the net cooling power and the minimum achievable temperature is investigated here. Three representative sky conditions will be considered: $e_{avg,sw} = 0.13$ at $T_a = 290\ K$ for the dry climates[43], $e_{avg,sw} = 0.53$ at $T_a = 300\ K$ for equatorial tropical climates[29] and $e_{avg,sw} = 0.33$ at $T_a = 295\ K$ for intermediate, mid-latitude climates. Two different degrees of thermal insulation, which determines the parasitic heat gain will be considered: $h_{eff} = 0$ and $h_{eff} = 10\ Wm^{-2}K^{-1}$. $h_{eff} = 10\ Wm^{-2}K^{-1}$ is a typical value readily achievable at ambient conditions by disposing an infrared-transmitting cover to prevent the convective heat gain[48] and an enclosing structure to reduce the radiative and conductive heat gains from the surrounding[18]. $h_{eff} = 0$ represents a limit where the conductive and convective parasitic heat gains are completely suppressed from the radiative cooling surface. The degree of thermal insulation approaching this limit can be obtained using a high vacuum chamber[16, 29, 49, 50]. For example, with a vacuum chamber pumped down to $10^{-5}$ torr, the estimated heat transfer coefficient $h_{eff} = 0.2 - 0.3 \, Wm^{-2}K^{-1}$ was demonstrated[16, 29].

*6.1 Net cooling power as a function of temperature reduction*

Fig. 2 shows the net cooling power $P_{rad}(T_a, T_s)$ of the blackbody emitter according to equation (6), in black lines, and the selective emitter according to equation (7), in blue lines, plotted as a function of the temperature reduction of the radiative cooling surface from the ambient temperature, $\Delta T = T_a - T_s$. The solid lines, $\eta_{max} = \frac{\pi}{2}$ or 90°, correspond to the case where the radiative cooling surface has a hemispherical access to the sky and no aperture mirror is used. The dotted lines are for $\eta_{max} = \frac{\pi}{4}$ or 45°, a representative opening angle for an aperture mirror. Figs. 2(a) to 2(c) correspond to the cases where $h_{eff} = 0$ and Figs. 2(d) to 2(f) correspond to the cases where $h_{eff} = 10 \, Wm^{-2}K^{-1}$. Figs. 2(a) and 2(d) are for dry climates, 2(b) and 2(e) for intermediate climates, and 2(c) and 2(f) for tropical climates.

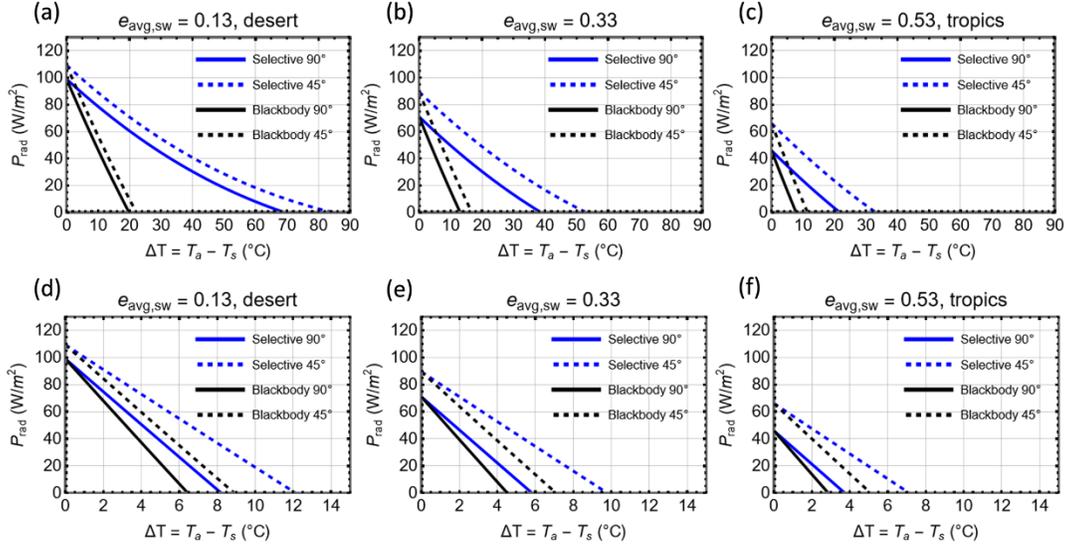

**Fig. 2** The net cooling power as a function of the temperature reduction from ambient temperature. The black lines and the blue lines represent the blackbody emitter and the selective emitter, respectively. 45° represents the opening angle of an aperture mirror. 90° corresponds to the case of no aperture mirror. Figs. 2(a) to 2(c) represent the case with no parasitic heat gain and Figs. 2(d) to 2(f) represent the case with finite parasitic heat gain.

The y-axis intercept of each curve corresponds to the cooling power, which is the net cooling power $P_{rad}$ at ambient temperature, $T_a = T_s$. This definition of the cooling power is the commonly used in the literature. The cooling power according to this definition corresponds to, for example, the instantaneous radiative cooling power at the onset of the radiative cooling effect or the radiative cooling power of a surface thermalised to the ambient. For clarity, the radiative cooling power according to equations (2), (6) and (7) at a temperature other than the ambient temperature, $T_a \neq T_s$, will be referred to as the "net cooling power" to avoid confusion with the conventionally defined "cooling power." The x-axis intercept is the maximum temperature reduction from the ambient temperature at the steady-state minimum temperature $\Delta T = T_a - T_{s\_min}$, which occurs when $P_{rad}(T_a, T_{s\_min}) = 0$. $T_{s\_min}$ represents the minimum temperature achievable with no cooling load. Figs. 2(a) to 2(f) clearly show that, the cooling power and the maximum temperature drop decrease as the average sky window emissivity $e_{avg,sw}$ decreases. This suggests that one and the same radiative cooling device will provide more cooling power and a higher temperature reduction in a dry climate than in a humid climate. Therefore, the performance of a radiative cooling surface is intricately coupled to the regional climate and the weather conditions. This evident point is often overlooked in the current literature, where the performance of a radiative cooling substrate is evaluated under certain atmospheric conditions and reported as if due only to the intrinsic properties of the substrate[51].

The maximum temperature drop, the x-axis intercept, of the selective emitter is larger than that of the blackbody, as is well understood[52]. At a given temperature below ambient, the selective emitter is always superior to the blackbody emitter in terms of the net cooling power because there is warming contribution outside the sky window on the blackbody emitter below the ambient temperature. Importantly, given the same conditions, namely the same sky emissivity and the same aperture mirror geometry, the cooling power, the y-axis intercept, is the same for the selective emitter and the blackbody emitter. This would not be the case if the "reciprocal ray" term $P_3$ is not considered. Similar analyses without the inclusion of the "reciprocal ray" term $P_3$ in ref [43] predict that the cooling power of blackbody is larger than that of the selective emitter and that there can be a cross-over point temperature before which the net cooling power of the blackbody emitter is larger than that of the selective emitter. The inclusion of $P_3$ renders the cooling power of the blackbody emitter and the selective emitter the same. This is deemed physically correct considering that, for both emitters, there is zero net thermal radiation for the wavelength band outside the sky window: the blackbody emitter, at ambient temperature, emits the same amount of thermal radiation as the incoming atmospheric radiation and the selective emitter fully reflects the incoming atmospheric radiation. Therefore, the net gain in cooling power occurs only within the sky window and is equal for both blackbody emitter and selective emitter.

As observed by comparing Figs. 2(a) to 2(c) with Figs. 2(d) to 2(f), respectively, the temperature reduction is enhanced by improving the thermal insulation while the cooling power, the y-axis intercept, is not affected. The enhancement of the temperature drop is most effective in the dry climate, where for the selective emitter with a mirror aperture with 45° opening angle, $\Delta T$ is increased by a factor of 9.3, reaching 80 °$C$. This suggests that in the arid regions, a substantial net cooling power can be obtained even at far below the freezing temperature and the dew point temperature, which is often sub-zero in such regions. A zero-energy refrigeration device or a water-harvesting device can be envisioned using a radiative cooling device aided by a vacuum shield.

*6.2 Cooling power as a function of opening angle*

It has been established so far that, under the same sky conditions and for the same mirror geometry, the cooling power is independent of whether the emitter is blackbody or selective and of the degree of thermal insulation. The cooling power itself, however, can be enhanced by an aperture mirror. Figs. 3(a) to 3(c) show the plots of the cooling power as a function of the opening angle of the aperture mirror $\eta_{max}$, under three different sky conditions. $\eta_{max} = 0$ corresponds to the case of no aperture mirror being used and $\eta_{max} = \frac{\pi}{2}$ or 90°, corresponds to the limit where the aperture mirror is sufficiently deep such that the radiative cooling surface is exposed to the sky only in the zenith direction. The aperture mirror in this limit is not practical to manufacture and deploy and considered here to be a reference point for the maximum obtainable power. In practice, the opening angle $\eta_{max}$ can range from 40° to 60° to provide an enhancement effect close to but smaller than this limit, as shown in Figs. 3(a) to 3(c). The use of the aperture mirror enhances the cooling power for all three sky conditions, however, each to a different degree. The maximum enhancement obtainable, given by the ratio of the cooling power at $\eta_{max} = \frac{\pi}{2}$ to that at $\eta_{max} = 0$, are 10.4 %, 26.0 % and 43.9 %, for the desert, the intermediate and the tropics, respectively. The absolute increment of cooling power from $\eta_{max} = \frac{\pi}{2}$ to $\eta_{max} = 0$, are $11.6\ W/m^2$, $21.5\ W/m^2$ and $24.3\ W/m^2$, for the desert, the intermediate and the tropics, respectively. This suggests that the use of the aperture mirror is most effective in the tropical climate for enhancing the cooling power.

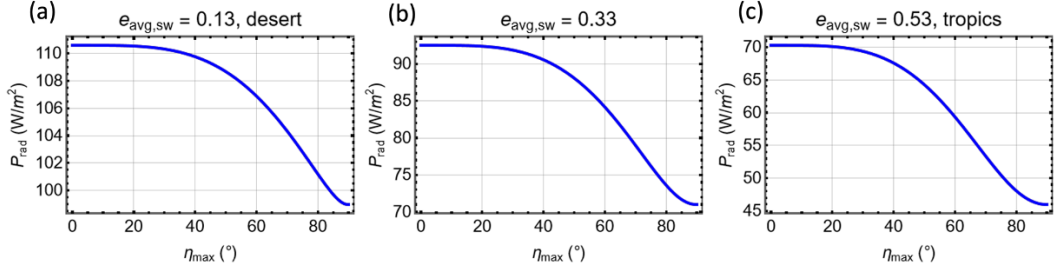

**Fig. 3** The cooling power of the blackbody emitter and the selective emitter as a function of the opening angle of the aperture mirror under three different sky conditions. As discussed in the text, the cooling power of the blackbody emitter and the selective emitter are the same under the same sky condition. 0° represents the limit where the aperture mirror exposes the radiative cooling surface only in the zenith direction. 90° corresponds to the case of no aperture mirror. Figs. 3(a) to 3(c) represent the sky conditions for the dry climate, the intermediate climate, and the humid climate, respectively.

*6.3 Maximum achievable temperature reduction as a function of opening angle*

Figs. 4(a) to 4(d) show the plots of the maximum achievable temperature reduction as a function of the opening angle of the aperture mirror $\eta_{max}$, under three different sky conditions. Figs. 4(a) and 4(b) correspond to the cases of the selective emitter and the blackbody emitter, respectively, in vacuum where $h_{eff} = 0$ and Figs. 4(c) and 4(d) correspond to the cases of the selective emitter and the blackbody emitter, respectively, where $h_{eff} = 10\ Wm^{-2}K^{-1}$. For the selective emitter in vacuum, in Fig. 4(a), the maximum enhancement obtainable, given by the ratio of the temperature reduction at $\eta_{max} = \frac{\pi}{2}$ to that at $\eta_{max} = 0$, are 25.3 %, 44.0 % and 64.3 %, for the desert, the intermediate and the tropics, respectively. The absolute increment in the temperature reduction are 17.5 °C, 17.1 °C and 14.0 °C, respectively. For the blackbody emitter in vacuum, in Fig. 4(b), the maximum enhancement obtainable is 13.3 %, 33.2 % and 56.3 %, respectively. The absolute increment in the temperature reduction are 2.6 °C, 4.3 °C and 4.4 °C, respectively. For the selective emitter with $h_{eff} = 10\ Wm^{-2}K^{-1}$, in Fig. 4(c), the maximum enhancement obtainable is 11.9 %, 30.5 % and 53.2 %, respectively. The absolute increment in the temperature reduction are 1.0 °C, 1.8 °C and 2.0 °C, respectively. For the blackbody emitter with $h_{eff} = 10\ Wm^{-2}K^{-1}$, in Fig. 4(d), the maximum enhancement obtainable is 11.9 %, 30.6 % and 53.4 %, respectively. The absolute increment in the temperature reduction are 0.8 °C, 1.4 °C and 1.5 °C, respectively. With the exception of the selective emitter in vacuum in terms of the absolute increment in the temperature reduction, the use of aperture mirror was most effective in the tropical climate in all other cases. Therefore, a similar conclusion can be drawn as the cooling power, namely that the use of the aperture mirror is most effective in the tropical climate for enhancing the temperature reduction.

In summary, our analyses with the blackbody emitter and the selective emitter show that the performance of an aperture mirror-enhanced radiative cooling system exhibits a clear dependence on the regional climate. By including the "reciprocal ray" contribution in the net cooling power, the physical requirement is met that the cooling power at ambient temperature is independent of the emissivity outside the sky window. Given the same conditions, namely the same sky window emissivity and the same aperture mirror geometry, the temperature reduction by the selective emitter is always larger than that of the blackbody emitter but the cooling power is the same for both emitters. Improving the degree of thermal insulation enhances the temperature reduction but not the cooling power. The enhancement of temperature reduction obtained by suppressing parasitic heat gain is the largest in the dry climate, where for the selective emitter in vacuum with a mirror aperture with 45° opening angle, the temperature reduction from the ambient temperature is over 80 °$C$. Rather counterintuitively, the use of an aperture mirror structure is most effective in the tropical climate, both in terms of the cooling power and the temperature reduction. In particular, the cooling power enhancement in the tropical climate can be as large as 43.9 %. Therefore, the use of an aperture mirror can be a useful augmentation to aid in overcoming the emissive sky of the tropical climate.

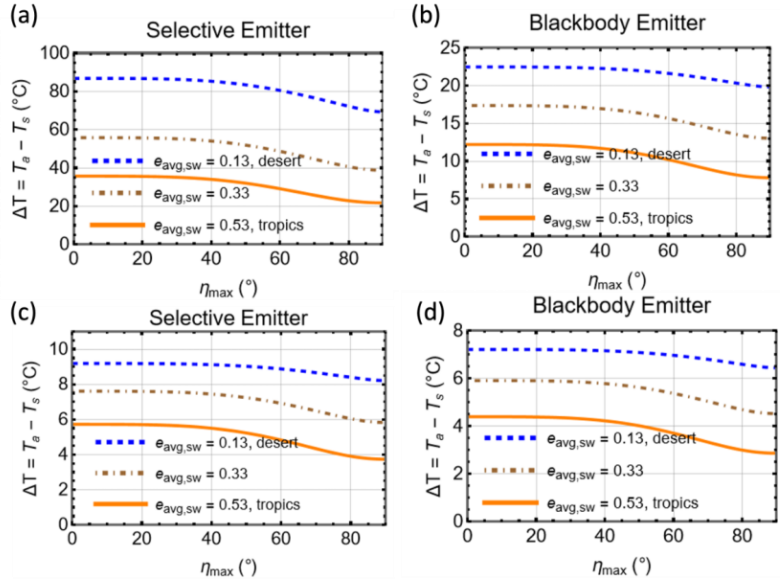

**Fig. 4** The maximum achievable temperature reduction of the selective emitter and the blackbody emitter as a function of the opening angle of the aperture mirror under three different sky conditions. 0º represents the limit where the aperture mirror exposes the radiative cooling surface only in the zenith direction. 90º corresponds to the case of no aperture mirror. Figs. 4(a) and 4(b) represent the case with no parasitic heat gain and Figs. 4(c) and 4(d) represent the case with a finite parasitic heat gain.

## 7  Regional dependence of enhancement of a realistic emitter: 3M specular reflector

For the blackbody emitter and the selective emitter according to equations (6) and (7), the emissivity spectrum $e_s(\lambda)$ was assumed to be angle-independent. However, the emissivity spectrum $e_s(\eta,\lambda)$ of most realistic radiative cooling surface depends on both angle $\eta$ and wavelength $\lambda^{53}$. In this case, the atmospheric model according to equation (1) can be applied to evaluate the net cooling power according to equation (2). As an example of a realistic radiative cooling surface, the 3M™ Enhanced Specular Reflector film (ESR)[54], referred to here as the specular reflector, will be considered. A mass-produced product with known optical properties[55, 56], it was demonstrated as an effective radiative cooling substrate[15, 57]. The angle-dependent emissivity spectrum $e_s(\eta, \lambda)$ of the specular reflector published in ref[15] will be used for numerical evaluation. The hemispherical emissivity is $\varepsilon = 0.60$ at $T_a = 300$ K according to $\varepsilon = \int_0^{\pi/2} d(sin^2\eta) \int_0^\infty d\lambda P(\lambda, T_a) e_s(\eta, \lambda) / \sigma T_a^4$. The emissivity within the sky window is $\varepsilon_{SW} = 0.79$ at $T_a = 300$ K according to $\varepsilon_{SW} = \int_0^{\pi/2} d(sin^2\eta) \int_{SW} d\lambda P(\lambda, T_a) e_s(\eta, \lambda) / \int_{SW} d\lambda P(\lambda, T_a)$.

*7.1 Net cooling power as a function of temperature reduction*

Fig. 5 shows the net cooling power $P_{rad}(T_a, T_s)$ of the specular reflector according to equation (2) plotted as a function of the temperature reduction of the radiative cooling surface from the ambient temperature, $\Delta T = T_a - T_s$. Figs. 5(a) to 5(c) correspond to dry climates, intermediate and tropical climates, respectively. The solid lines correspond to the case of no aperture mirror, $\eta_{max} = \frac{\pi}{2}$ or 90°. The dotted lines correspond to an aperture mirror with $\eta_{max} = \frac{\pi}{4}$ or 45°. The blue lines represent the case with no parasitic heat gain, $h_{eff} = 0$ and the black lines correspond to the cases with $h_{eff} = 10 \, Wm^{-2}K^{-1}$. Many of the observations from the blackbody emitter the selective emitter examples in section 6 hold also for the specular reflector. For example, the cooling power and the maximum temperature drop decrease with the increasing average sky window emissivity $e_{avg,sw}$. The enhancement of the temperature drop, by suppressing the parasitic heat gain, is most effective in the dry climate, where, with a mirror aperture with 45° opening angle, $\Delta T$ is increased by a factor of 5, reaching over 33 °C.

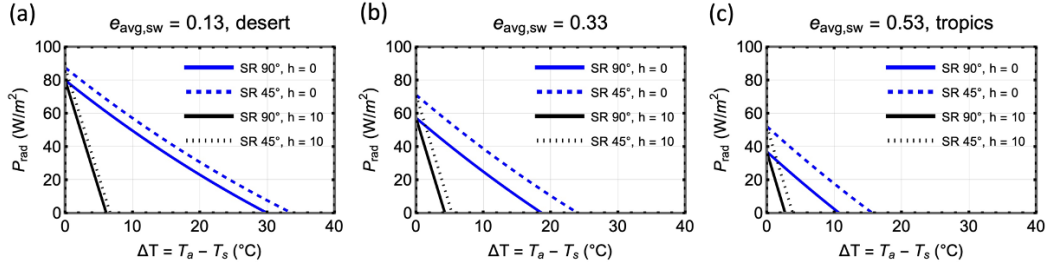

**Fig. 5** The net cooling power of the 3M specular reflector as a function of the temperature reduction from ambient temperature. The black lines the case with finite parasitic heat gain and the blue lines represent the case with no parasitic heat gain. The dotted lines represent 45º opening angle of an aperture mirror. The solid lines represent 90º opening angle or no aperture mirror. Figs. 5(a) to 5(c) correspond to the dry climate, the intermediate climate and the humid climate, respectively.

It is mentioned in passing that the heat gain due to the solar absorption is omitted for simplicity in the net cooling power according to equations (2), (6) and (7). Since the recent demonstration of the daytime radiative cooing under direct sunlight[14, 15], radiative cooling has attracted a broader interest. The key technical challenge for the daytime radiative cooling is to prevent the heat gain from solar absorption by reflecting the incident solar irradiation. There have been considerable efforts to optimize the radiative cooling surface with an ultimate end-goal of a unity reflectivity in the solar spectrum[58, 59] and recently a solar reflectivity of 99.6 % has been demonstrated[60]. Any solar heat gain would shift the curves of Figs. 2 and 5 to in the negative y-direction, thereby reducing the cooling power and the maximum temperature reduction. For example, in Fig. 5(c), for the case of the specular reflector without any aperture mirror or vacuum thermal insulation, represented by the black solid curve, a realistic 3 % absorption of the irradiation of one sun, 1000 $W/m^2$ would reduce the cooling power from 36 $W/m^2$ to 6 $W/m^2$ and reduce the achievable temperature reduction from 2.7 °$C$ to 0.5 °$C$. Considering that other imperfections and compromising factors can be present, such as the cloud coverage, the observation of sub-ambient temperature during daytime remains a challenge in the equatorial tropical climate[29, 36, 61, 62] or even in tropical and humid sub-tropical climates[39, 63-68]. Fig. 5(c) hints that by improving the thermal insulation with a high vacuum chamber and the use of an aperture mirror structure, sub-ambient radiative cooling is possible, as was demonstrated during the daytime under a cloudy sky in Singapore[29].

*7.2 Cooling power as a function of opening angle*

Figs. 6(a) to 6(c) show the plots of the cooling power as a function of the opening angle of the aperture mirror $\eta_{max}$, under three different sky conditions. The opening angle $\eta_{max}$ ranges from 15° to 85° as in the data used for the emissivity spectrum $e_s(\eta, \lambda)$[15]. The maximum enhancement obtainable, given by the ratio of the cooling power at $\eta_{max} = 85°$ to that at $\eta_{max} = 15°$, are 11.2 %, 29.6 % and 52.7 %, for the desert, the intermediate and the tropics, respectively. The absolute increment of cooling power from $\eta_{max} = 85°$ to $\eta_{max} = 15°$, are 8.9 $W/m^2$, 17.0 $W/m^2$ and 19.5 $W/m^2$, respectively. This shows that, also for the specular reflector, the use of the aperture mirror is most effective in the tropical climate for enhancing the cooling power.

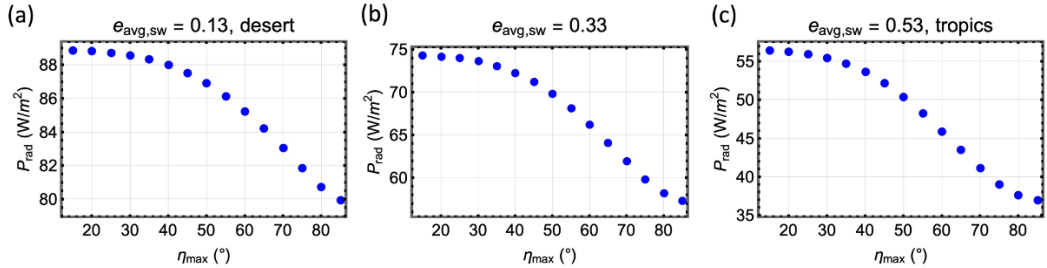

**Fig. 6** The cooling power of the 3M specular reflector as a function of the opening angle of the aperture mirror under three different sky conditions. 0° represents the limit where the aperture mirror exposes the radiative cooling surface only in the zenith direction. 90° corresponds to the case of no aperture mirror. Figs. 6(a) to 6(c) represent the sky conditions for the dry climate, the intermediate climate, and the humid climate, respectively.

*7.3 Maximum achievable temperature reduction as a function of opening angle*

Figs. 7(a) and 7(b) show the plots of the maximum achievable temperature reduction as a function of the opening angle of the aperture mirror $\eta_{max}$, under three different sky conditions. Fig. 7(a) corresponds to the cases in vacuum where $h_{eff} = 0$ and Fig. 7(b) corresponds to the cases where $h_{eff} = 10\ Wm^{-2}K^{-1}$. For the specular reflector in vacuum, in Fig. 7(a), the maximum enhancement obtainable, given by the ratio of the temperature reduction at $\eta_{max} = 85°$ to that at $\eta_{max} = 15°$, are 14.4 %, 35.2 % and 58.8 %, for the desert, the intermediate and the tropics, respectively. The absolute increment in the temperature reduction are 4.3 °C, 6.6 °C and 6.3 °C, respectively. For the specular reflector with $h_{eff} = 10\ Wm^{-2}K^{-1}$, in Fig. 7(b), the maximum enhancement obtainable, given by the ratio of the temperature reduction at $\eta_{max} = 85°$ to that at $\eta_{max} = 15°$, are 11.3 %, 29.9 % and 53.1 %, for the desert, the intermediate and the tropics, respectively. The absolute increment in the temperature reduction are 0.7 °C, 1.3 °C and 1.4 °C, respectively. The aperture mirror is in general effective in the tropical climate in terms for enhancing the temperature reduction although a clear conclusion can be drawn only for the case with $h_{eff} = 10\ Wm^{-2}K^{-1}$.

To summarise, the 3M specular reflector with the average emissivity of 0.60 and the emissivity within the sky window of 0.79 was applied to the model atmosphere as an example of a realistic emitter surface. Many conclusions from the analyses of the blackbody emitter and the selective emitter example also hold: the cooling power and the maximum temperature reduction decrease with a higher average sky window emissivity and improving the thermal insulation enhances the temperature reduction but not the cooling power. Also as in the blackbody emitter and the selective emitter, the enhancement of temperature reduction obtained by suppressing parasitic heat gain is the largest in the dry climate. For the specular reflector in vacuum with a mirror aperture with 45° opening angle, the temperature reduction from the ambient temperature is over 33 °C. For the realistic emitter also, the use of an aperture mirror structure is most effective in the tropical climate, both in terms of the cooling power and the temperature reduction. In particular, the cooling power enhancement in the tropical climate can be as large as 52.7 %, amplified from 37.0 $W/m^2$ to 56.5 $W/m^2$. The daytime radiative cooling in the equatorial tropical climate remains a challenge, if not impractical, due to the highly emissive sky and the solar trajectory through the zenith. The analyses presented herein shed light on the strategy, namely the combination of the thermal insulation using a high vacuum and the aperture mirror structure.

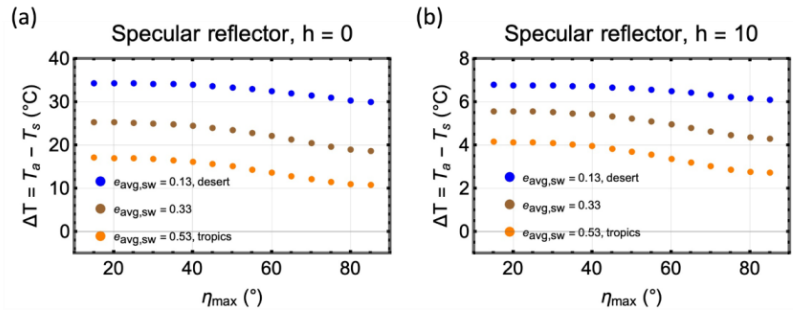

**Fig. 7** The maximum achievable temperature reduction of the 3M specular reflector as a function of the opening angle of the aperture mirror under three different sky conditions. 0° represents the limit where the aperture mirror exposes the radiative cooling surface only in the zenith direction. 90° corresponds to the case of no aperture mirror. Fig. 7(a) represents the case with no parasitic heat gain and Fig. 7(b) represents the case with a finite parasitic heat gain.

## 8  Real-time monitoring of the average sky window emissivity

The atmospheric spectrum can be characterised according to equation (1) with the knowledge of two parameters: the ambient temperature $T_a$ and the average sky window emissivity $e_{avg,sw}$. The net cooling power according to equation (2) and the steady state temperature, $T_s$ when $P_{rad}(T_a, T_s) = 0$, can be evaluated with the knowledge of one more parameter: the angle-dependent emissivity spectrum of the radiative cooling substrate $e_s(\eta, \lambda)$.

The change in time of the atmospheric model according to equation (1) can be obtained by measuring the average sky window emissivity $e_{avg,sw}$ in real time. This can be achieved by measuring the thermal radiation power within the sky window wavelength range, within a small solid angle in the zenith direction[29], for example, with an infrared thermometer[47] with the active range limited to be within the sky window and with a small field of view. Many infrared thermometers are equipped with a bandpass filter with a passband overlapping the sky window for the long-range applications. For example, an IR pyrocouple (Calex PC151LT-0mA) is commercially available with an active response in the 8 to 14 $\mu m$ wavelength range, a fixed emissivity of $\varepsilon_{IR} = 0.95$ and a field of view 15:1. The infrared power received by such IR thermometer corresponds to the $P_2$ of equation (4) with $\eta_{max} = 1/15$ and $e_{s,sw}(\eta, \lambda) = \varepsilon_{IR} = 0.95$ and $e_{s,nsw}(\eta, \lambda) = 0$.

$$P_{IRTherm} = \varepsilon_{IR} \int_0^{\eta_{max}} d(\sin^2 \eta) \int_0^{\infty} d\lambda P(\lambda, T_a) e_a(\eta, \lambda) e_s(\eta, \lambda)$$
$$= \varepsilon_{IR} \int_0^{\eta_{max}} d(\sin^2 \eta) \int_{SW} d\lambda P(\lambda, T_a) \left(1 - (1 - e_{avg,sw})^{\frac{1}{\cos \eta}}\right) \propto \varepsilon_{IR} \int_{SW} d\lambda P(\lambda, T_a) \, e_{avg,sw}. \quad (11)$$

The last approximation is based on that when $\eta_{max} = 1/15$, $\int_0^{\eta_{max}} d(\sin^2 \eta) \sim \eta_{max}^2$ and $1/\cos \eta_{max} \sim 1$. The power detected by the infrared thermometer is therefore directly proportional to the sky window emissivity in the zenith direction, $e_{avg,sw}$. In practice, the IR thermometer directly outputs a signal corresponding to the sky window temperature $T_{SW}$, which is calibrated to the received infrared power within the 8 to 14 $\mu m$ wavelength range of $\varepsilon_{IR} P(\lambda, T_{SW})$. The sky window emissivity in the zenith direction, $e_{avg,sw}$ can be obtained from the sky window temperature $T_{SW}$ with $e_{avg,sw} = \int_{SW} d\lambda \varepsilon_{IR} P(\lambda, T_{SW}) / \int_{SW} P(\lambda, T_a) d\lambda$.

Fig. 8 illustrates the relationship between the average sky window emissivity $e_{avg,sw}$ and the sky window temperature $T_{SW}$. The example relates to the MODTRAN simulation results[41] with the pre-set locality of "Tropical Atmosphere" with the following parameters: 'CO2 400ppm, CH4 1.7ppm, Troposphere Ozone 28 ppb, Altitude 0 km looking up, Stratosphere Ozone scale 1, Water vapour scale 1, Freon scale 1, No clouds or Rain.' The purple solid line is the MODTRAN simulation result for the atmospheric radiation towards the earth at ambient temperature $T_a$ = 299.7 K. The Planck spectrum at the ambient temperature is shown as the black dotted curve. Outside the sky window of 8 to 14 $\mu m$ wavelength range, the MODTRAN atmospheric spectrum closely follows the Planck spectrum. The ratio of the atmospheric emission to the Planck spectrum is shown as the purple solid line in Fig. 8(b). As discussed in the equation (11), the infrared thermometer measures the emission power over the sky window and outputs a corresponding temperature. For the atmosphere represented by the MODTRAN simulation results presented in Fig. 8(a), the infrared thermometer pointing towards the zenith direction would measure -8.5 °C, in other words, the sky window temperature, $T_{SW}$, of 264.5 K. The Planck spectrum at this temperature is shown as the blue dotted line in the left panel. The atmospheric model according to equation (1) constructed therefrom is shown as the red solid line, both in Figs. 8(a) and 8(b). In Fig. 8(a), the atmospheric emission spectrum follows the Planck spectrum at the ambient temperature $P(\lambda, T_a)$ outside the sky window, and the Planck spectrum at the sky window temperature $P(\lambda, T_{SW})$ within the sky window. In Fig. 8(b), the emissivity spectrum of the atmosphere is $e_{avg,sw}$, a single measured value within the sky window and unity outside the sky window. The average sky window emissivity $e_{avg,sw}$ according to the MODTRAN simulation is 0.52 as shown in the right panel of Fig. 8(b). The sky window temperature, $T_{SW}$, is to be differentiated with the conventionally defined "sky temperature," $T_{sky}$, defined as $P_{Pyrgeo} = \sigma T_{sky}^4$, where $P_{Pyrgeo}$ is the atmospheric emission over the hemisphere as measured with a Pyrgeometer. The power measured by a Pyrgeometer corresponds to $P_2$ of equation (4) with the hemispherical acceptance $\eta_{max} = \pi/2$, assuming the receiving substrate to be a black body, $e_s(\eta, \lambda) = 1$.

$$P_{Pyrgeo} = \int_0^{\pi/2} d(sin^2\eta) \int_0^\infty d\lambda P(\lambda, T_a) e_a(\eta, \lambda) e_s(\eta, \lambda) = \int_0^{\pi/2} d(sin^2\eta) \int_0^\infty d\lambda P(\lambda, T_a) e_a(\eta, \lambda)$$

$$= \int_0^{\pi/2} d(sin^2\eta) \left( \int_{SW} d\lambda P(\lambda, T_a) \left(1 - (1 - e_{avg,sw})^{\frac{1}{\cos\eta}}\right) + \int_{NSW} d\lambda P(\lambda, T_a) \right)$$

$$= \sigma T_a^4 - \int_0^{\pi/2} d(sin^2\eta) \left( \int_{SW} d\lambda P(\lambda, T_a)(1 - e_{avg,sw})^{\frac{1}{\cos\eta}} \right). \tag{12}$$

The thermopile inside a Pyrgeometer measures a quantity corresponding to the second term of the last line of equation (12). The second term also represents the long wave imbalance, the net difference between the blackbody emission at the ambient temperature $T_a$ and the atmospheric emission.

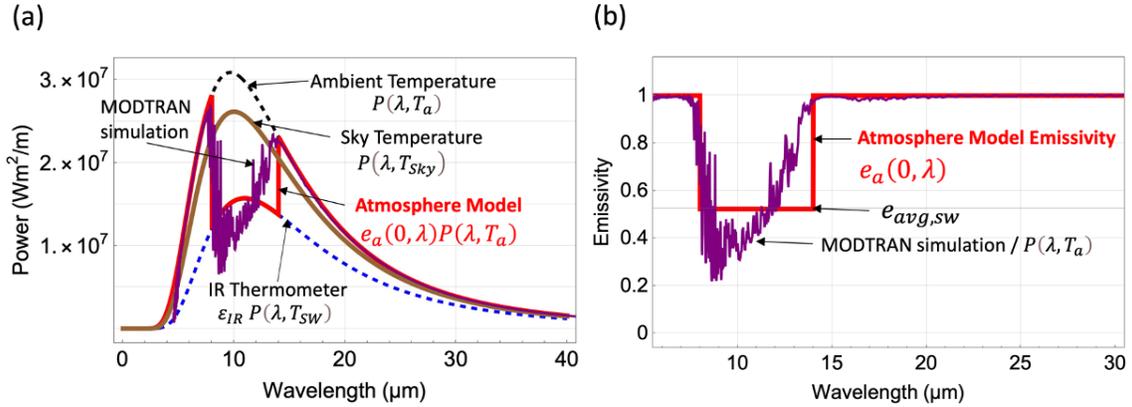

**Fig. 8** Illustration of the atmospheric model and the method to measure the average sky window emissivity using an infrared thermometer, in reference to the Planck spectrum and a simulated atmospheric spectrum. (a) the MODTRAN simulation of atmospheric spectrum of the tropical atmosphere (purple) and the model atmosphere (red). The Planck spectrum at the ambient temperature (black dotted), the sky temperature (brown) and the sky window temperature (blue dotted) are shown. (b) the emissivity spectrum of the atmosphere model (red) shown with the simulation result normalized with the Planck spectrum at the ambient temperature (purple).

Fig. 8(a) illustrates the relationship between the sky window temperature $T_{SW}$ according to the atmospheric model of equation (1) and the conventionally defined sky temperature $T_{sky}$. Assuming a uniform sky, the total hemispherical atmospheric emission $P_{Pyrgeo}$ and the sky temperature $T_{sky}$ evaluated based on this atmospheric model according to equation (12) is $T_{sky}$= 289.9 K. The Planck spectrum at the sky temperature $P(\lambda, T_{sky})$ is shown in the left panel as a brown solid line. It is noted that the definition of the sky temperature, $P_{Pyrgeo} = \sigma T_{sky}^4$, assumes that the overall atmospheric emission follows a Planck spectrum with the total power $P_{Pyrgeo}$. Therefore, the area under the atmospheric spectrum, either from the MODTRAN simulations (purple solid line) or from the atmospheric model of equation (1) (red solid line), equals the area under the Planck spectrum at the sky temperature $P(\lambda, T_{sky})$. As a result, the ambient temperature $T_a$ is higher than both the sky temperature $T_{sky}$ and the sky window temperature $T_{SW}$. The sky window temperature $T_{SW}$ is lower than the sky temperature $T_{sky}$.

It is to be noted that the range of the sky window in the example of Fig. 8 was set to be 8 to 14 $\mu m$ wavelength range to match the active range of the example infrared thermometer, whereas for the results presented in Figs. 2 to 7, the sky window was set to be 7.9 to 13 $\mu m$ wavelength range for direct comparison with ref [43]. The exact range of the sky window may slightly differ in each report but these differences cause only a minor adjustment of the average sky window emissivity $e_{avg,sw}$ and do not change the area under the atmospheric spectrum, represented in Fig. 8(a) as the red, purple and brown lines. Therefore, the same results are obtained in terms of the performance measures such as the net cooling power and the minimum achievable temperature.

If the sky condition is uniform throughout the hemisphere, the average sky window emissivity $e_{avg,sw}$ can be estimated with the measurement of the sky temperature, $T_{sky}$ and vice versa. However, when the sky condition is not uniform, any irregular cloud pattern, especially the presence of the low-lying cloud patches, would increase the discrepancy between the two measurements. This is because the Pyrgeometer measurement corresponds to the integration over the hemisphere and over the whole infrared spectrum, so the spectral and angular information are obscured, while the infrared thermometer measures in the zenith direction and within the sky window. For example, the Pyrgeometer measurement would not differentiate between a sky with a single, thick cloud patch blocking the zenith direction and a sky with a widely distributed thin clouds, which would lead to very different radiative cooling performances. The atmospheric model according to equation (1), combined with the measurement with an infrared thermometer discussed here can be a way of a realistic characterization of the sky condition for radiative cooling. Especially for the radiative cooling devices with an aperture mirror structure described herein, the direct measurement of the average sky window emissivity $e_{avg,sw}$ can provide a more accurate evaluation of the relevant sky condition because it is responsive only within the sky window and a small solid angle around the zenith. For example, it is known that radiative cooling surface warms up when the cloud screens the clear sky[16, 47, 69] but no detailed studies are available on the cloud base temperature. The direct measurement infrared power within a small solid angle around the zenith can provide the cloud base temperature, which varies depending on the cloud height, thickness and layer compositions[70, 71]. When the clouds are constantly moving in and out of the field of view of the radiative cooling setup, the cloud base temperature measured by the infrared thermometer can be taken to be the effective sky window temperature $T_{SW}$ and the sky window emissivity $e_{avg,sw}$ is evaluated also for the cloud base[29].

## 9  Conclusion

The net cooling power of an aperture mirror-enhanced radiative cooling substrate was investigated. The dependence on various parameters have been explored, such as the sky window emissivity, the opening angle of the aperture mirror and the degree of thermal insulation. The cooling power and the temperature reduction are the highest in a dry climate, but the aperture mirror structure is most effective in a humid climate.

Many currently proposed radiative cooling surfaces for the large-area radiative cooling are in the form of a paint, a polymeric film or a ceramic plate and additional measures are rarely considered for the geometrical amplification or for the fortified thermal isolation. This is motivated mainly by the scalability, the affordability, and the convenience in handling of the radiative cooling material. The efforts for optimisation are often directed to the development of the radiative cooling surface itself, mainly the improvement of the solar reflectivity for daytime radiative cooling. However, the unity emissivity and unity solar reflectivity are not the ultimate end goal: even after perfecting these spectral properties, the cooling power and the temperature reduction can be pushed further using enhancing measures. As shown in the analyses presented here, there still exists a large room for improvement in the parameter space for the geometrical amplification and for the thermal insulation. Although the additional structures for the fortified thermal insulation and the geometrical amplification would add to the cost and the complexity, these are not fundamental limitations and should be viewed as the new avenue of engineering challenges. For example, the high-vacuum packaging is common in the field of solar thermal harvesting and the aperture mirror structures can be mass-produced with the help of 3-D printing.

The enhancement obtained by an aperture mirror structure can be critical for practical applications, for example, building structures in that a higher cooling power can lower the target temperature and shorten the time constant for a given thermal load. Therefore, using the results presented in this work, a systemic approach can be taken in the design of the radiative cooling device including reflector structures and thermal insulation layers, considering the local climate.

*Code and Data Availability Statement*

All data in support of the findings of this paper are available within the article.

*Acknowledgments*

This research is supported by the National Research Foundation, Singapore and A*STAR under its CQT Bridging Grant. The author thanks Prof. Christian Kurtsiefer for hosting him at the Centre for Quantum Technologies at NUS and Dr. Yu-Hung Lien for fruitful discussions.

*References*

1. P. Nema, S. Nema and P. Roy, "An overview of global climate changing in current scenario and mitigation action," *Renewable & Sustainable Energy Reviews* **16** (4), 2329-2336 (2012). https://doi.org/10.1016/j.rser.2012.01.044.
2. Y. Xu, V. Ramanathan and D. G. Victor, "Global warming will happen faster than we think," Nature Publishing Group UK London (2018).
3. I. P. o. C. Change, "Global warming of 1.5° C," *World Meteorological Organization: Geneva, Switzerland*, (2018).
4. I. E. Agency, "The Future of Cooling: Opportunities for energy- efficient air conditioning," [https://webstore.iea.org/the-future-of-cooling]
5. H. E. Landsberg, *The urban climate*, Academic press (1981).
6. F. Trombe, "Perspectives sur l'utilisation des rayonnements solaires et terrestres dans certaines régions du monde. ," *Rev. Gén. Therm.* **6**, 1285–1314 (1967).
7. A. W. Harrison and M. R. Walton, "RADIATIVE COOLING OF TIO2 WHITE PAINT," *Solar Energy* **20** (2), 185-188 (1978). https://doi.org/10.1016/0038-092x(78)90195-0.
8. C. G. Granqvist, "RADIATIVE HEATING AND COOLING WITH SPECTRALLY SELECTIVE SURFACES," *Applied Optics* **20** (15), 2606-2615 (1981). https://doi.org/10.1364/ao.20.002606.
9. A. Hjortsberg and C. G. Granqvist, "RADIATIVE COOLING WITH SELECTIVELY EMITTING ETHYLENE GAS," *Applied Physics Letters* **39** (6), 507-509 (1981). https://doi.org/10.1063/1.92783.
10. T. S. Eriksson and C. G. Granqvist, "RADIATIVE COOLING COMPUTED FOR MODEL ATMOSPHERES," *Applied Optics* **21** (23), 4381-4388 (1982). https://doi.org/10.1364/ao.21.004381.

11. E. M. Lushiku, A. Hjortsberg and C. G. Granqvist, "RADIATIVE COOLING WITH SELECTIVELY INFRARED-EMITTING AMMONIA GAS," *Journal of Applied Physics* **53** (8), 5526-5530 (1982). https://doi.org/10.1063/1.331487.
12. P. Berdahl, M. Martin and F. Sakkal, "THERMAL PERFORMANCE OF RADIATIVE COOLING PANELS," *International Journal of Heat and Mass Transfer* **26** (6), 871-880 (1983). https://doi.org/10.1016/s0017-9310(83)80111-2.
13. A. R. Gentle and G. B. Smith, "Radiative Heat Pumping from the Earth Using Surface Phonon Resonant Nanoparticles," *Nano Letters* **10** (2), 373-379 (2010). https://doi.org/10.1021/nl903271d.
14. A. P. Raman et al., "Passive radiative cooling below ambient air temperature under direct sunlight," *Nature* **515** (7528), 540-+ (2014). https://doi.org/10.1038/nature13883.
15. A. R. Gentle and G. B. Smith, "A Subambient Open Roof Surface under the Mid-Summer Sun," *Advanced Science* **2** (9), (2015). https://doi.org/10.1002/advs.201500119.
16. Z. Chen et al., "Radiative cooling to deep sub-freezing temperatures through a 24-h day-night cycle," *Nature Communications* **7**, (2016). https://doi.org/10.1038/ncomms13729.
17. J. L. Kou et al., "Daytime Radiative Cooling Using Near-Black Infrared Emitters," *Acs Photonics* **4** (3), 626-630 (2017). https://doi.org/10.1021/acsphotonics.6b00991.
18. B. Bhatia et al., "Passive directional sub-ambient daytime radiative cooling," *Nature Communications* **9**, (2018). https://doi.org/10.1038/s41467-018-07293-9.
19. D. L. Zhao et al., "Subambient Cooling of Water: Toward Real-World Applications of Daytime Radiative Cooling," *Joule* **3** (1), 111-123 (2019). https://doi.org/10.1016/j.joule.2018.10.006.
20. L. Zhou et al., "A polydimethylsiloxane-coated metal structure for all-day radiative cooling," *Nature Sustainability* **2** (8), 718-724 (2019).
21. W. Li and S. Fan, "Radiative cooling: harvesting the coldness of the universe," *Optics and Photonics News* **30** (11), 32-39 (2019).
22. M. M. Hossain and M. Gu, "Radiative cooling: principles, progress, and potentials," *Advanced Science* **3** (7), 1500360 (2016).
23. E. E. Bell et al., "Spectral radiance of sky and terrain at wavelengths between 1 and 20 microns. II. Sky measurements," *JOSA* **50** (12), 1313-1320 (1960).
24. L. Zhou et al., "Hybrid concentrated radiative cooling and solar heating in a single system," *Cell Reports Physical Science* **2** (2), (2021).
25. J. Peoples et al., "Concentrated radiative cooling," *Applied Energy* **310**, 118368 (2022).
26. M. Sheng et al., "Characterization and performance enhancement of radiative cooling on circular surfaces," *Renewable and Sustainable Energy Reviews* **188**, 113782 (2023).
27. Z. Chen et al., "Simultaneously and synergistically harvest energy from the sun and outer space," *Joule* **3** (1), 101-110 (2019).
28. I. Haechler et al., "Exploiting radiative cooling for uninterrupted 24-hour water harvesting from the atmosphere," *Science Advances* **7** (26), eabf3978 (2021).
29. J. Hwang, "Daytime radiative cooling under extreme weather conditions," *arXiv preprint arXiv:2310.09304*, (2023).
30. M. Dong et al., "Concentrated radiative cooling and its constraint from reciprocity," *Optics Express* **30** (1), 275-285 (2022).
31. A. Gentle et al., "3D printable optical structures for sub-ambient sky cooling," *Thermal Radiation Management for Energy Applications* 16-24 (2017).

32. J. Liu et al., "Recent advances in the development of radiative sky cooling inspired from solar thermal harvesting," *Nano Energy* **81**, 105611 (2021).
33. A. Gentle, K. Dybdal and G. Smith, "Polymeric mesh for durable infra-red transparent convection shields: applications in cool roofs and sky cooling," *Solar energy materials and solar cells* **115**, 79-85 (2013).
34. A. Gentle, J. Aguilar and G. Smith, "Optimized cool roofs: Integrating albedo and thermal emittance with R-value," *Solar Energy Materials and Solar Cells* **95** (12), 3207-3215 (2011).
35. C. Liu et al., "Effect of atmospheric water vapor on radiative cooling performance of different surfaces," *Solar Energy* **183**, 218-225 (2019).
36. D. Han, B. F. Ng and M. P. Wan, "Preliminary study of passive radiative cooling under Singapore's tropical climate," *Solar Energy Materials and Solar Cells* **206**, (2020). https://doi.org/10.1016/j.solmat.2019.110270.
37. X. Yu and C. Chen, "A simulation study for comparing the cooling performance of different daytime radiative cooling materials," *Solar Energy Materials and Solar Cells* **209**, 110459 (2020).
38. J. Feng et al., "Thermal analysis in daytime radiative cooling," *IOP Conference Series: Materials Science and Engineering* 072064 (2019).
39. M. Dong et al., "Nighttime radiative cooling in hot and humid climates," *Optics express* **27** (22), 31587-31598 (2019).
40. P. Berdahl, "Retrospective on the resource for radiative cooling," *Journal of Photonics for Energy* **11** (4), 042106-042106 (2021).
41. D. Archer, "MODTRAN, http://climatemodels.uchicago.edu/modtran/," University of Chicago, [http://climatemodels.uchicago.edu/modtran/]
42. A. Berk et al., "MODTRAN (R) 6: A major upgrade of the MODTRAN (R) radiative transfer code," in Proceedings of SPIE, *20th SPIE Conference on Algorithms and Technologies for Multispectral, Hyperspectral, and Ultraspectral Imagery* (2014), https://doi.org/10.1117/12.2050433.
43. G. B. Smith, "Amplified radiative cooling via optimised combinations of aperture geometry and spectral emittance profiles of surfaces and the atmosphere," *Solar Energy Materials and Solar Cells* **93** (9), 1696-1701 (2009). https://doi.org/10.1016/j.solmat.2009.05.015.
44. M. Martin and P. Berdahl, "Summary of results from the spectral and angular sky radiation measurement program," *Solar Energy* **33** (3-4), 241-252 (1984).
45. C. Granqvist and A. Hjortsberg, "Surfaces for radiative cooling: Silicon monoxide films on aluminum," *Applied Physics Letters* **36** (2), 139-141 (1980).
46. J. Feng et al., "Dynamic impact of climate on the performance of daytime radiative cooling materials," *Solar Energy Materials and Solar Cells* **208**, (2020). https://doi.org/10.1016/j.solmat.2020.110426.
47. J. L. C. Aguilar et al., "A method to measure total atmospheric long-wave down-welling radiation using a low cost infrared thermometer tilted to the vertical," *Energy* **81**, 233-244 (2015).
48. J. Zhang et al., "Cover shields for sub-ambient radiative cooling: A literature review," *Renewable & Sustainable Energy Reviews* **143**, (2021). https://doi.org/10.1016/j.rser.2021.110959.

49. X. Ao et al., "Self-adaptive integration of photothermal and radiative cooling for continuous energy harvesting from the sun and outer space," *Proceedings of the National Academy of Sciences* **119** (17), e2120557119 (2022).
50. U. Banik et al., "Impact of parasitic heat fluxes on deep sub-ambient radiative coolers under variable pressure," *Applied Thermal Engineering* **237**, 121655 (2024).
51. L. Zhou, X. Yin and Q. Gan, "Best practices for radiative cooling," *Nature Sustainability* **6** (9), 1030-1032 (2023).
52. S. Catalanotti et al., "The radiative cooling of selective surfaces," *Solar Energy* **17** (2), 83-89 (1975).
53. A. Gentle and G. Smith, "Angular selectivity: impact on optimised coatings for night sky radiative cooling," *Nanostructured Thin Films II* 96-103 (2009).
54. M. F. Weber et al., "Giant birefringent optics in multilayer polymer mirrors," *Science* **287** (5462), 2451-2456 (2000).
55. "Perkin Elmer Field Application Report, July 2023 https://resources.perkinelmer.com/lab-solutions/resources/docs/far_measurement-of-enhanced-specular-reflector-films-using-lambda-1050-and-ura-accessory-012190_01.pdf," (accessed July 2023 [https://resources.perkinelmer.com/lab-solutions/resources/docs/far_measurement-of-enhanced-specular-reflector-films-using-lambda-1050-and-ura-accessory-012190_01.pdf]
56. M. Janecek, "Reflectivity spectra for commonly used reflectors," *IEEE Transactions on Nuclear Science* **59** (3), 490-497 (2012).
57. A. Leroy et al., "High-performance subambient radiative cooling enabled by optically selective and thermally insulating polyethylene aerogel," *Science advances* **5** (10), eaat9480 (2019).
58. X. X. Yu, J. Q. Chan and C. Chen, "Review of radiative cooling materials: Performance evaluation and design approaches," *Nano Energy* **88**, (2021). https://doi.org/10.1016/j.nanoen.2021.106259.
59. M. J. Chen et al., "Passive daytime radiative cooling: Fundamentals, material designs, and applications," *Ecomat* **4** (1), (2022). https://doi.org/10.1002/eom2.12153.
60. K. Lin et al., "Hierarchically structured passive radiative cooling ceramic with high solar reflectivity," *Science* **382** (6671), 691-697 (2023).
61. D. Han et al., "The criteria to achieving sub-ambient radiative cooling and its limits in tropical daytime," *Building and Environment* **221**, (2022). https://doi.org/10.1016/j.buildenv.2022.109281.
62. D. Han et al., "Sub-ambient radiative cooling under tropical climate using highly reflective polymeric coating," *Solar Energy Materials and Solar Cells* **240**, 111723 (2022).
63. S. Y. Jeong et al., "Field investigation of a photonic multi-layered TiO2 passive radiative cooler in sub-tropical climate," *Renewable Energy* **146**, 44-55 (2020). https://doi.org/10.1016/j.renene.2019.06.119.
64. S. Y. Jeong et al., "Daytime passive radiative cooling by ultra emissive bio-inspired polymeric surface," *Solar Energy Materials and Solar Cells* **206**, (2020). https://doi.org/10.1016/j.solmat.2019.110296.
65. C. Y. Tso, K. C. Chan and C. Y. Chao, "A field investigation of passive radiative cooling under Hong Kong's climate," *Renewable energy* **106**, 52-61 (2017).
66. R. Y. Wong et al., "Critical sky temperatures for passive radiative cooling," *Renewable Energy* **211**, 214-226 (2023).

67. H. Zhong et al., "Highly solar-reflective structures for daytime radiative cooling under high humidity," *ACS Applied Materials & Interfaces* **12** (46), 51409-51417 (2020).
68. J. Liu et al., "Field investigation and performance evaluation of sub-ambient radiative cooling in low latitude seaside," *Renewable Energy* **155**, 90-99 (2020).
69. A. Gentle and G. Smith, "Performance comparisons of sky window spectral selective and high emittance radiant cooling systems under varying atmospheric conditions," *Proceedings Solar2010, The 48th AuSES Annual Conference* (2010).
70. A. Maghrabi, R. Clay and D. Riordan, "Detecting cloud with a simple infra-red sensor," *Transactions of the Royal Society of South Australia* **133** (1), 164-171 (2009).
71. D. Riordan et al., "Cloud base temperature measurements using a simple longwave infrared cloud detection system," *Journal of Geophysical Research: Atmospheres* **110** (D3), (2005).

**Caption List**

**Fig. 1** The geometry of aperture mirror structure.

**Fig. 2** The net cooling power as a function of the temperature reduction from ambient temperature.

**Fig. 3** The cooling power of the blackbody emitter and the selective emitter as a function of the opening angle of the aperture mirror under three different sky conditions.

**Fig. 4** The maximum achievable temperature reduction of the selective emitter and the blackbody emitter as a function of the opening angle of the aperture mirror under three different sky conditions.

**Fig. 5** The net cooling power of the 3M specular reflector as a function of the temperature reduction from ambient temperature.

**Fig. 6** The cooling power of the 3M specular reflector as a function of the opening angle of the aperture mirror under three different sky conditions.

**Fig. 7** The maximum achievable temperature reduction of the 3M specular reflector as a function of the opening angle of the aperture mirror under three different sky conditions.

**Fig. 8** Illustration of the atmospheric model and the method to measure the average sky window emissivity using an infrared thermometer, in reference to the Planck spectrum and a simulated atmospheric spectrum.